%% file: ms_clean.tex
\documentclass[a4paper,usenatbib,letters]{mnras}
\usepackage{graphicx}
\usepackage{amsmath}
\usepackage{siunitx}
\usepackage{hyperref}

\usepackage{bm}

\usepackage[T1]{fontenc}
\usepackage{ae,aecompl}
\usepackage{times}

\usepackage{multirow}

\usepackage{newtxtext,newtxmath}

\newcommand{\unitflux}{\,erg\,cm$^{-2}$\,s$^{-1}$}
\newcommand{\unitlumi}{\,erg\,s$^{-1}$}
\newcommand{\mission}[1]{\textit{#1}}

\newcommand{\lxray}{L_\mathrm{X}}

\title[ULX candidates in NGC 7090]{X-ray properties of two transient ULX candidates in galaxy NGC 7090}

\author[Z. Liu]{Zhu Liu,$^{1,2}$\thanks{Contact e-mail: \href{mailto:liuzhu@nao.cas.cn}{liuzhu@nao.cas.cn}}
    P. T. O'Brien,$^{2}$
    J. P. Osborne,$^{2}$
    P. A. Evans,$^{2}$
    K. L. Page,$^{2}$
\\
$^{1}$ Key Laboratory of Space Astronomy and Technology, National Astronomical Observatories, Chinese Academy of Sciences, Beijing 100012, China \\
$^{2}$ Department of Physics and Astronomy, University of Leicester, Leicester, UK}

\date{}
\pubyear{2018}

\begin{document}
    
\label{firstpage}
\pagerange{\pageref{firstpage}–\pageref{lastpage}}
\maketitle

\begin{abstract}
We report the X-ray data analysis of two transient ultraluminous X-ray sources (ULXs, hereafter X1 and X2) located in the nearby galaxy NGC 7090. While they were not detected in the 2004 \mission{XMM-Newton} and 2005 \mission{Chandra} observations, their 0.3--10\,keV X-ray luminosities reached $>3\times10^{39}$\unitlumi in later \mission{XMM-Newton} or \mission{Swift} observations, showing increases in flux by a factor of $>80$ and $>300$ for X1 and X2, respectively. X1 showed indications of spectral variability: at the highest luminosity, its X-ray spectra can be fitted with a powerlaw ($\Gamma=1.55\pm0.15$), or a multicolour disc model with $T_{\mathrm{in}}=2.07^{+0.30}_{-0.23}$\,keV; the X-ray spectrum became softer ($\Gamma=2.67^{+0.69}_{-0.64}$), or cooler ($T_\mathrm{in}=0.64^{+0.28}_{-0.17}$\,keV) at lower luminosity. No strong evidence for spectral variability was found for X2. Its X-ray spectra can be fitted with a simple powerlaw model ($\Gamma=1.61^{+0.55}_{-0.50}$), or a multicolour disc model ($1.69^{+1.17}_{-0.48}$\,keV). A possible optical counterpart for X1 is revealed in \mission{HST} imaging. No optical variability is found, indicating that the optical radiation may be dominated by the companion star. Future X-ray and optical observations are necessary to determine the true nature of the compact object. 
\end{abstract}


\section{Introduction}

Ultraluminous X-ray sources (ULXs) are point-like off-nuclear extragalactic sources with X-ray luminosity higher than $\sim10^{39}$\unitlumi \citep{fabbiano89, feng_soria11}. The apparent X-ray luminosity of ULXs exceeds the Eddington limit of a stellar black hole (BH) with a typical mass of $\sim10M_\odot$ found in Galactic BH X-ray binaries (BHXRBs, \citealt{remillard_mcclintock06}). It has been generally believed that ULXs are powered either by super-Eddington accretion onto stellar-mass black holes, or by intermediate mass black holes (IMBHs) with sub-Eddington accretion rate \citep[e.g.][]{colbert_mushotzky99, feng_soria11}. Observational evidence for stellar-mass BHs have been found in a few ULXs (e.g. M101 ULX-1, \citealt{liu_etal13}), while ESO 243-49 HLX1 \citep{farrell_etal09} and M82 X-1 \citep{feng_kaaret10, pasham_etal14}, both with relatively high peak X-ray luminosity ($\lxray\geq10^{41}$\,\unitlumi), are promising IMBH candidates. However, the detection of pulsations in the X-ray data of four ULXs (M82 X-2: \citealt{bachetti_etal14}; NCG 5907 ULX-1: \citealt{israel_etal17b}; NGC 7793 P13: \citealt{furst_etal16, israel_etal17a}; NGC 300 ULX-1: \citealt{carpano_etal18}) show clear evidence that the accretors in those systems are neutron stars (NS), indicating that the apparent X-ray luminosities in those ULXs are at least $\ge10$ times the Eddington limit for a standard NS of mass $1.4M_\odot$.

\begin{figure*}
\begin{center}
\includegraphics[width=1.0\textwidth]{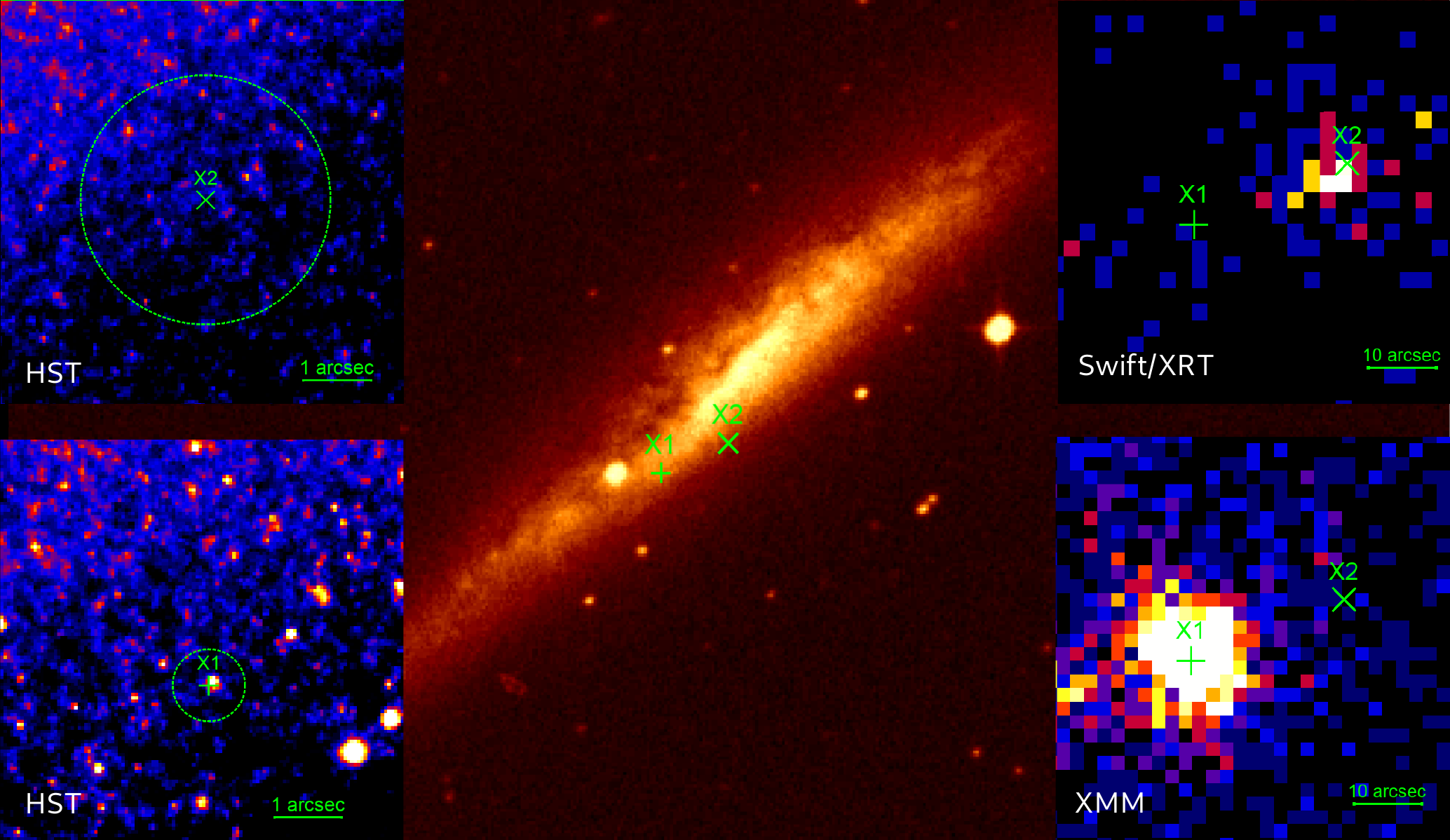}
\vspace*{-5mm}
\caption{\label{fig:srcimg}\textit{Main image}: DSS image of NGC 7090. The plus and cross symbols mark the position of X1 (measured from \mission{Chandra}) and X2 (measured from \mission{Swift/XRT}), respectively. \textit{Bottom-left} subset: \mission{HST} F814W image in the region around X1. The circle indicates the 90 per cent \mission{Chandra} position uncertainty (0.54\,arcsec) of X1. \textit{Top-left} subset: \mission{HST} F804W image in the region around X2. The position uncertainty of X2 is 1.8\,arcsec (dashed circle, 90 per cent confidence level). \textit{Top-right} subset: combined (observations from 4 June to 2 July 2012) \mission{Swift/XRT} image. X2 is clearly detected by \mission{XRT}. \textit{Bottom-right} subset: mosaic \mission{XMM-Newton} EPIC image. The position of X1 measured from \mission{Chandra} is consistent with the source detected by \mission{XMM-Newton}.}
\end{center}
\end{figure*}

Some ULXs show low level short-term variability with fractional variability $\ll 10$ per cent, while some may be highly variable with fractional variability $\sim10-30\,$ per cent \citep[e.g.][]{heil_etal09, sutton_etal13, middleton_etal15}. ULXs with long-term flux variability by a factor of $\sim40-1000$, though quite rare, have also been found, e.g. NGC 3628 \citep{strickland_etal01}, M101 ULX-X1 \citep{mukai_etal05}, M82 X2 \citep{feng_kaaret07}, NGC 1365 ULX X2 \citep{soria_etal09}, CXOM31 J004253.1+411422 \citep{kaur_etal12} and XMMU J004243.6+412519 \citep{esposito_etal13} in M31 and NGC 5907 ULX-2 \citep{pintore_etal18}. All the four pulsar ULXs discovered so far are also highly variable (even transient) X-ray sources.

The X-ray spectra of many luminous ULXs ($L_\mathrm{X}>3\times10^{39}$\unitlumi) can generally be fitted with either a two component model (the ultraluminous state, UL), i.e., a multicolour disc blackbody (DBB) plus a Comptonisation or a single Comptonisation component \citep{gladstone_etal09, sutton_etal13}. For the less luminous ULXs ($L_\mathrm{X}<3\times10^{39}$\unitlumi), their spectra can be well described with a single $p$-free disc model (the broadened disk, BD; \citealt{sutton_etal13}) for which the local disc temperature $T(r)$ is proportional to $r^{-p}$. Some ULXs with luminosity higher than $10^{40}$\unitlumi also show a spectral shape consistent with the BD model \citealt{pintore_etal16}. Spectral variability has been revealed in some individual ULXs through detailed X-ray spectral or colour analysis \citep[e.g.][]{kubota_etal01, roberts_etal06, feng_kaaret09, kajava_poutanen09}. Some ULXs, similar to the Galactic X-ray binaries (XRBs), can change their spectral state dramatically \citep{sutton_etal13, marlowe_etal14}, e.g. Holmberg IX X-1 showed a two component disc plus power-law spectrum at lower luminosity, while the spectral shape changed to a broadened disc at higher luminosity \citep{walton_etal14, luangtip_etal16}. The spectral properties of the four pulsar ULXs are similar to typical ULXs, although pulsar ULXs show a further excess at high energy whose origin may be associated to the accretion column above the NS surface. However, even though less robustly, indications of such an excess are observed also in other non-pulsating ULXs, suggesting that the ULX population can host a larger number of neutron stars than previously expected \citep{walton_etal18a}. 

In this letter, we report the X-ray properties of two transient ULXs (Fig.\,\ref{fig:srcimg}, hereafter X1, X2) found in the nearby star-forming galaxy NGC 7090. X1 is classified as an ULX candidate in \citet{lin_etal12} based on \mission{XMM-Newton} observations. X2 was detected in the 2012 \mission{Swift/XRT} observations and included in the first \mission{Swift}-XRT point source catalogue \citep[1SXPS][]{evans_etal14}. In this work we identify it as an ULX with peak 0.3--10\,keV X-ray luminosity higher than $3\times10^{39}$\unitlumi. We adopted a distance to NGC 7090 of $6.6$\,Mpc \citep{tully_etal92} throughout this work .

\section{Data analysis}

NGC 7090 was observed by \mission{XMM-Newton}, \mission{Chandra}, \mission{Swift} and \mission{Hubble} in the past decades. The observation log can be found in Table\,\ref{tab:obs_log}. In this section, we describe the details of the data analysis.

\input{obs_log}

\subsection{\mission{XMM-Newton}}

NGC 7090 was observed by \mission{XMM-Newton} on 2004 April 18 (ObsID: 0200230101), 2004 May 13 (ObsID: 0200230201) and 2007 October 5 (ObsID: 0503460101) with exposure time 28ks, 19ks and 31ks, respectively. The first observation was severely affected by high background flaring, and thus was excluded from this work. The \mission{XMM-Newton} Science Analysis System (\textsc{SAS}) software version 16.1 \citep{gabriel_etal04} was used to reduce \mission{XMM-Newton} data. We ran \textsc{SAS} tasks \textsc{emchain} and \textsc{epchain} to generate the event lists for the European Photon Imaging Camera (EPIC) MOS \citep{turner_etal01} and pn \citep{struder_etal01} detectors, respectively. Flaring background periods were identified and filtered from the event lists. The effective exposure time of the EPIC pn, M1 and M2 cameras, after filtering the high background periods, were 6024, 10390, 10290\,s (4270, 12790, 12980\,s) for the 2001 (2007) observation, respectively. Source detection was performed on all the individual EPIC image as well as the combined EPIC image for each observation using the \textsc{SAS} task \textsc{edetect\_chain}. The parameters \textit{likemin} (minimum detection likelihood) and \textit{mlmin} (minimum likelihood) of 8 and 10 were adopted as suggested by the \textsc{SAS} guide\footnote{\url{https://xmm-tools.cosmos.esa.int/external/sas/current/doc/eboxdetect/node3.html}}. We found X2 was not detected in the individual EPIC image or in the combined image of the two observations, while X1 was only detected in the October 2007 observation (both in the individual image and the combined image). We thus only extracted the X-ray spectra for source X1. A circular region with a radius of 12\,arcsec was used to extract the source spectra. Apart from X2, we note that X1 is also about 19\,arcsec away from the closest source, and it is $\sim5$ times brighter than that source during the 2007 \mission{XMM-Newton} observation. X-ray events with pattern $\le 12$ and $\le 4$ were selected to extract the MOS and pn spectrum, respectively. The background spectra were extracted from a source-free region with a circle of radius 100\,arcsec located on the same CCD chip as the source for MOS, while a circular region centred at the same CCD read-out column as the source position was selected for pn. The \textsc{arfgen} and \textsc{rmfgen} tasks were used to generate the response files.

\subsection{\mission{Chandra}}

\mission{Chandra} observed NGC 7090 on 2005 December 18 (26ks, ObsID: 7060) and 2006 April 10 (31ks, ObsID: 7250) with the Advanced CCD Imaging Spectrometer (ACIS). \mission{Chandra} data were reduced with \textsc{CIAO} (\citealt{fruscione_etal06}, ver 4.10) software package and calibration files CALDB (ver 4.7.6). We ran \textsc{wavdetect} tool on the \mission{Chandra} observations to generate a source list. X2 was not detected in the two \mission{Chandra} observations, while X1 was only detected in the 2006 observation. The overall 90 per cent absolute astrometry uncertainty of \mission{Chandra} is $\sim0.8$\,arcsec\footnote{\url{http://cxc.cfa.harvard.edu/cal/ASPECT/celmon}}. Following the online data analysis guide\footnote{\url{http://cxc.harvard.edu/ciao/threads/reproject_aspect}}, we corrected the absolute astrometry by cross-matching \mission{Chandra} sources with the GAIA DR2 catalogue \citep{gaiadr2} using a correlation radius of 1\,arcsec. Three sources were selected to perform absolute astrometry correction. The \textsc{CIAO} task \textsc{wcs\_match} and \textsc{wcs\_update} were used to correct and update the aspect ratio. The residual rms scatter in the corrected X-ray positions of the GAIA sources is 0.26\,arcsec, which corresponds to a 90 per cent position error of $\approx0.53$\,arcsec (assuming Rayleigh distribution).

To extract the source spectrum for X1, we selected a circular region with a radius of 2\,arcsec. The background spectrum was extracted from an annulus (concentric with the source) region with an inner and outer radius of 6 and 24\,arcsec, respectively. The regions surrounding the events from the X2 (a circle with radius of 2.7\,arcsec) and the other close by source (a circle with radius of 5.2\,arcsec) were excluded from the annulus background region. The \textsc{CIAO} task \textsc{dmextract} was used to extract the source and background spectra. The response files are generated using the \textsc{mkacisrmf} and \textsc{mkarf} tasks.

\subsection{\mission{Swift/XRT} observations}

NGC 7090 was observed by the X-ray Telescope (\mission{XRT}, \citealt{burrows_etal05}) of the \mission{Neil Gehrels Swift Observatory} (\mission{Swift}) from 2006 to 2014. All the XRT data (21 observations) were analysed with the XRT online data analysis tool\footnote{\url{http://www.swift.ac.uk/user_objects}}\citep{evans_etal09}. We ran source detection using $\textsc{ximage}$ task $detect$. Source X1 was not detected in either individual observations or the combined observation with signal-to-noise ratio (S/N) more than 2, while source X2 was detected in observations performed on 2012 June 4 and July 2 (ObsIDs: 00032287008, 00032287011) with S/N higher than 3.7 \citep{evans_etal14}. To increase the S/N, we extracted the source and background spectra from a combined image of the four observations performed between 2012 June 4 and July 2 (ObsIDs: 00032287008-11, total exposure: 11\,ks, $\mathrm{S/N}>9$, hereafter \mission{Swift}1). X2 was also detected in the combined image of observations performed from 2012 July 30 to August 20 (ObsIDs: 00032287011-15, total exposure: 8.7\,ks, $\mathrm{S/N}>5$, hereafter \mission{Swift}2). Source and background spectra were also extracted for this combined observation.

\subsection{\mission{HST}}

NGC 7090 was observed six times by \mission{HST} from 1994 to 2016. In this work, the observations with better S/N carried out on 2001 September 24 with the Wide Field and Planetary Camera 2 (WFPC2, filter $F814W$), on 2005 June 23 with the Wide Field Camera 3 (ACS/WFC3, filter $625W$) and on 2012 April 9 with ACS/WFC3 (filters: $F814W$ and $F606W$) were used\footnote{Note that the 2007 \mission{HST} observation had a very long exposure time. However, no photometric measurements were given on the \mission{Hubble} Source Catalogue website. Thus this observation is not used in this work.}. The \mission{HST} images were retrieved from the \mission{Hubble Legacy Archive}\footnote{\url{http://hla.stsci.edu}} (HLA). The absolute astrometry for the 2012 observations (which have the best spatial resolution and S/N) was corrected by aligning the \mission{HST} images with the source positions found in the GAIA DR2 catalogue. The absolute astrometry accuracy of \mission{HST} after correction is $\sim1\,\mathrm{mas}$ (68 per cent confidence level), consistent with the position accuracy obtained at HLA.

\section{Results}
The position of X1 was obtained from the 2006 \mission{Chandra} observation using the \textsc{wavdetect} task, which gives $\mathrm{RA}=21^\mathrm{h}36^\mathrm{m}31^\mathrm{s}.81$ and $\mathrm{Dec.}=\ang{-54;33;57.82}$, within the error circle of the position measured from the \mission{XMM-Newton} 2007 observation. Following \citet{evans_etal14}, we improved the position accuracy of \mission{Swift/XRT} by aligning the \mission{XRT} image with the sources detected by \mission{Chandra}. The improved position of X2 given by \mission{XRT} is then: $\mathrm{RA}=21^\mathrm{h}36^\mathrm{m}29^\mathrm{s}.11$ and $\mathrm{Dec.}=\ang{-54;33;48.31}$ (with 90 per cent uncertainty of $1.8\,\mathrm{arcsec}$), which is about 25.3\,arcsec away from X1.

\subsection{\label{xray_var}X-ray variability}

\begin{figure}
\begin{center}
\includegraphics[width=\columnwidth]{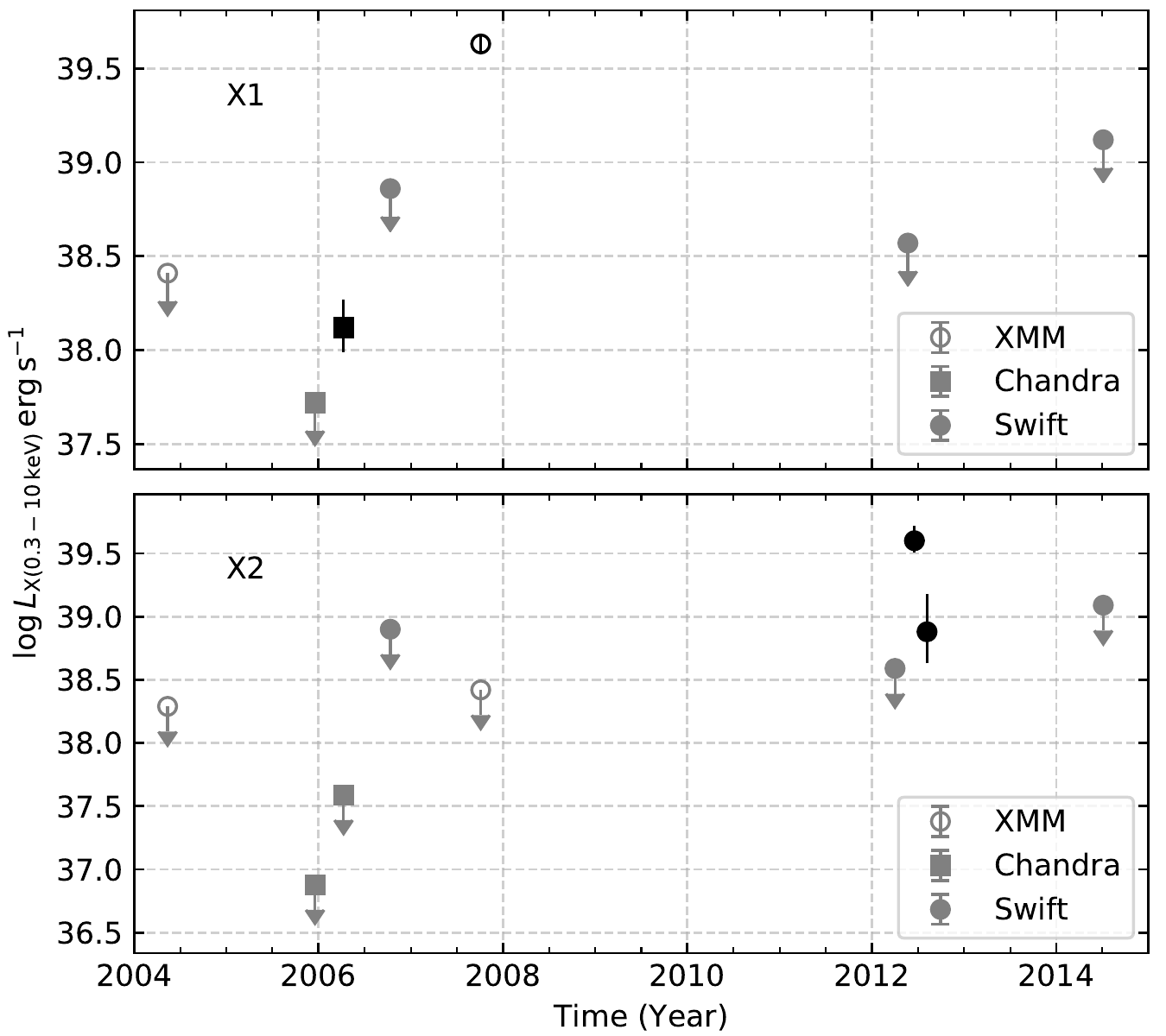}
\vspace*{-0.5cm}
\caption{\label{fig:lc}The long-term unabsorbed $0.3-10$\,keV light curves of X1 (top) and X2 (bottom). The errors on the luminosities are at 90 per cent confidence level, while the upper-limits (in grey) are at the $3\sigma$ confidence level.}
\end{center}
\end{figure}

Fig.~\ref{fig:lc} shows the long-term X-ray variability of the two ULXs. The unabsorbed 0.3--10\,keV X-ray luminosities of X1, estimated by fitting the X-ray spectra with an absorbed power-law model (see Sec.\,\ref{x1_spec} for more details), are $\lxray\sim1\times10^{38}$ and $\sim4\times10^{39}$\unitlumi for the 2006 April \mission{Chandra} and 2007 October \mission{XMM-Newton} observations, respectively. The $3\sigma$ upper limits, estimated using the best-fitting absorbed power-law model of the 2007 \mission{XMM-Newton} observation, are plotted for the other observations (or the \mission{Swift/XRT} combined observations) for which X1 was not detected. The lowest X-ray luminosity was given by the 2005 \mission{Chandra} observation with a $3\sigma$ upper limit of $\sim5\times10^{37}$\unitlumi.

Source X2 was significantly detected by \mission{Swift/XRT} in the observations made on 2012 June 4 ($>3\sigma$) and July 2 ($>5\sigma$). It was also seen in the other two observations performed in 2012 June, albeit with less significance ($\sim2.6\sigma$). The average X-ray luminosity estimated by fitting the average X-ray spectrum of those four observations with an absorbed power-law model is $\sim4\times10^{39}$\unitlumi (see Fig.~\ref{fig:lc}). X2 was not detected in any of the other individual observations. But it was clearly seen in the combined image of the \mission{Swift/XRT} data observed between 2012 July 30 and August 20 (\mission{Swift}2) with an estimated X-ray luminosity of $7\times10^{38}$\unitlumi (see Fig.~\ref{fig:lc}). The lowest X-ray luminosity was calculated from the 2006 \mission{Chandra} observation with a $3\sigma$ upper limit of $8\times10^{36}$\unitlumi.

From Fig.~\ref{fig:lc}, it is clear that both X1 and X2 showed dramatic long-term X-ray variability. Comparing to the 2005 \mission{Chandra} observation, the highest X-ray luminosity of X1 (the 2007 \mission{XMM-Newton} observation) and X2 (the 2012 \mission{Swift/XRT} observations) increased by a factor of $>80$ and $>300$, respectively. We also analysed the temporal properties of X1 within the 2007 \mission{XMM-Newton} data. No significant short-term (e.g. minutes to hours) variability was found in the 31\,ks exposure time. We did not find any coherent signal in the power spectrum created using the $0.3-10\,\mathrm{keV}$ \mission{XMM-Newton} data. Assuming a sinusoidal modulation, a $3\sigma$ upper limit of $\sim60$ per cent on the pulsed fraction (defined as the semi-amplitude of the sinusoid divided by the source average count rate) was derived using the \mission{XMM-Newton} data for periods in the range of $\sim0.4-150\,\mathrm{s}$.

\input{./table_fitting.tex}

\subsection{X-ray spectral analysis}
X-ray spectral analysis was carried out for the 2006 \mission{Chandra} (background subtracted 0.3--10\,keV photon counts $C_\mathrm{sub}=34$) and 2007 \mission{XMM-Newton} ($C_\mathrm{sub}=350$, 353 and 283 for EPIC M1, M2 and pn, respectively) observations of X1, as well as the two X-ray spectra of X2 ($C_\mathrm{sub}=110$ and 15 for \mission{Swift}1 and \mission{Swift}2, respectively). \textsc{Xspec} (\citealt{arnaud96} ver 12.10) is used to fit the X-ray spectra. The cash statistic (wstat in \textsc{Xspec}) is used due to the relatively low photon counts. Galactic and host galaxy absorption are included in all models (model \textsc{tbabs} and \textsc{ztbabs} in \textsc{Xspec}, abundances are set to \texttt{wilm}, \citealt{wilms_etal00}). The Galactic absorption is fixed at $5.4\times10^{20}$\,$\mathrm{cm}^{-2}$ \citep{kalberla_etal05}. Quoted uncertainties on spectral parameters are the 90 per cent confidence limits unless stated otherwise.

\subsubsection{\label{x1_spec}Source X1}
\begin{figure}
\begin{center}
\includegraphics[width=\columnwidth]{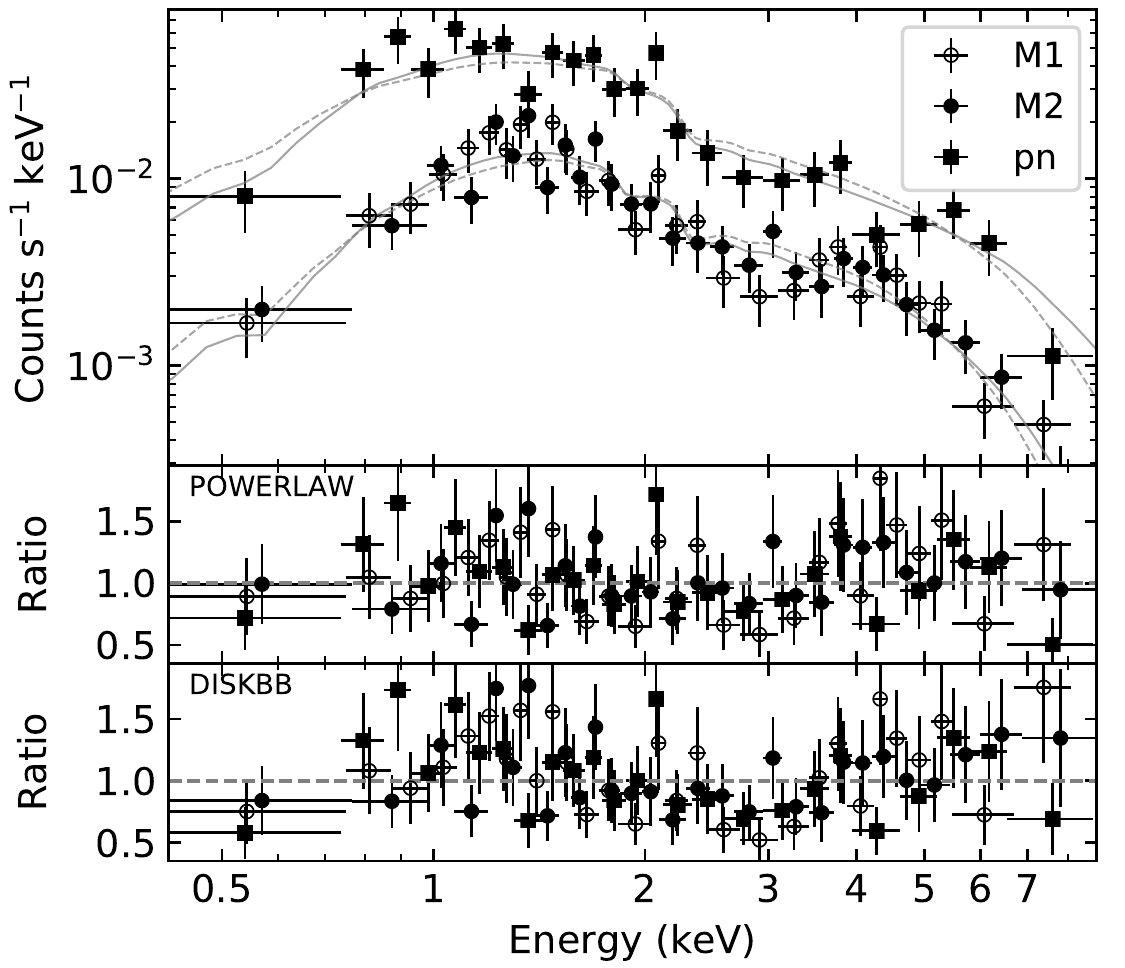}
\vspace*{-0.5cm}
\caption{\label{fig:spec}Top panel: the X-ray spectra of X1 during the 2007 observation: open circle for EPIC-M1, filled circle for EPIC-M2, square for EPIC-pn. The grey solid and dashed lines show the best-fitting \textsc{powerlaw} and \textsc{diskbb} model, respectively. The lower two panels show the data/model ratio for different models.}
\end{center}
\end{figure}

The EPIC 0.3--10\,keV M1, M2 and pn spectra were fitted simultaneously. A normalization factor is included to account for the calibration differences between the detectors. We fitted the data with two simple models: \textsc{powerlaw} model (\textsc{cons*tbabs*ztbabs*powerlaw} in \textsc{Xspec}) and \textsc{diskbb} model (\textsc{cons*tbabs*ztbabs*diskbb}). Both models can fit the data well (see Fig.\,\ref{fig:spec}) with $\Gamma = 1.55\pm0.15$ in the \textsc{powerlaw} model and inner disc temperature $T_\mathrm{in}=2.1^{+0.3}_{-0.2}$\,keV for \textsc{diskbb} model. Best-fitting values of the intrinsic absorption are $5.0^{+1.0}_{-1.0}\times10^{21}$\,$\mathrm{cm}^{-2}$ (\textsc{powerlaw}) and $3.0^{+1.0}_{-1.0}\times10^{21}$\,$\mathrm{cm}^{-2}$ (\textsc{diskbb}). The estimated unabsorbed 0.3-10\,keV X-ray luminosity is higher than $3\times10^{39}$\,\unitlumi. We also tried to fit the data with two component models, i.e. a \textsc{powerlaw} plus a \textsc{diskbb}, which gave $T_\mathrm{in}=0.22^{+0.18}_{-0.08}$\,keV and $\Gamma=1.44^{+0.26}_{-0.29}$, or a \textsc{blackbody} plus a \textsc{diskbb} ($T_\mathrm{BB}=1.41^{+0.24}_{-0.16}$\,keV, $T_\mathrm{in}=0.40^{+0.14}_{-0.09}$\,keV). Those two component models improve the fit slightly comparing with a single powerlaw model ($\Delta{C}=5.4$ and $11.3$ for 2 d.o.f, see Table\,\ref{tab:fit_para}). The best-fitting models as well as the data-to-model ratios for the \textsc{powerlaw} and \textsc{diskbb} models are shown in Fig.\,\ref{fig:spec}. The 2006 \mission{Chandra} data were fitted with the two simple models. The intrinsic column was fixed at values found by \mission{XMM-Newton} data. Although with large uncertainties, the data suggest a steeper photon index ($\Gamma=2.67^{+0.69}_{-0.64}$) or a lower disc temperature ($T_\mathrm{in}=0.64^{+0.28}_{-0.17}$\,keV), indicating a change in spectral shape. The best-fitting parameters of different models can be found in Table\,\ref{tab:fit_para}.

\subsubsection{\label{x2_spec}Source X2}
The same simple models were fitted to the \mission{Swift/XRT} spectra for X2. Both models can fit the \mission{Swift}1 spectrum well, with best-fitting temperature $T_\mathrm{in}=1.69^{+1.17}_{-0.48}$\,KeV or photon index $\Gamma=1.61^{+0.55}_{-0.50}$. Due to the low S/N, we did not fit the X-ray spectrum with more complicated models. The best-fitting results of the \mission{Swift}2 spectrum were consistent with the \mission{Swift}1 data, though with large uncertainties (intrinsic column was fixed at the values found by \mission{Swift}1) and relatively small change in flux (by a factor of $\sim4$).

\subsection{Optical counterpart}

We found one optical counterpart (see Fig.\,\ref{fig:srcimg}) within the position uncertainty of X1 in the \mission{HST} images. The AB magnitudes of the X1 counterpart (obtained from the \mission{Hubble} Source Catalogue) are: $23.27^{+0.06}_{-0.06}$, $23.32^{+0.01}_{-0.01}$, $23.28^{+0.02}_{-0.02}$ and $23.45^{+0.02}_{-0.02}\,\mathrm{mag}$ for the $WFPC2/F814W$ (2001), $ACS/F625W$ (2005), $ACS/F814W$ (2012) and $ACS/F606W$ (2012), respectively. Assuming $N_\mathrm{H}=2.21\times10^{21}A_\mathrm{V}$ \citep{guver_ozel09} and $A_V=E(B-V)/3.1$, the estimated extinction for $ACS/F814W$, $ACS/F625W$ and $ACS/F606W$ are $A_{F814W}=1.3$, $A_{F625W}=1.9$ and $A_{F606W}=2.1$ \citep{sirianni_etal05} with $N_\mathrm{H, host}=3\times10^{21}\,\mathrm{cm^{-2}}$, respectively. The estimated $V$, $R$ and $I$ band magnitudes, transformed from the $ACS/WFC$ AB magnitude \citep{sirianni_etal05}, are 21.28, 21.24 and 21.49\,mag, respectively. No significant variability is found for the $F814W$ flux in the two \mission{HST} observations.

Multiple optical counterparts were found within the position uncertainty of X2 in the \mission{HST} images. The magnitudes of the brightest source in the 2012 \mission{HST} observation were 24.86 and 25.93 mag in $F814W$ and $F606W$ band,\footnote{There is no photometric measurement for the 2005 observation on the \mission{Hubble} Source Catalogue.} respectively. Assuming $N_\mathrm{H, host}=3\times10^{21}\,\mathrm{cm^{-2}}$, the upper limit magnitudes for the X2 counterpart are $24.0$ and $23.1$ mag in the $V$ and $I$ bands, respectively.

\section{Discussion}

In this letter, we report the X-ray properties of two highly variable ULXs in the nearby star-forming galaxy NGC 7090. Source X1 has been classified as an ULX candidate in the catalogue compiled by \citet{lin_etal12} using \mission{XMM-Newton} data. Source X2 is a new ULX detected in the 2012 \mission{Swift/XRT} observations. The long-term X-ray light curves show that both sources are highly variable: flux changed by a factor of $>80$ for X1 and $> 300$ for X2. AGNs are known to be highly variable especially in the X-ray bands. However, variability by a factor of more than $80$ are rare in AGNs \citep[e.g.][]{strotjohann_etal16}. We further explore the possibility that the two sources are background AGNs by considering the $\log{N}-\log{S}$ of extragalactic X-ray sources. The expected number of AGN, with X-ray flux higher than $\sim10^{-13}$\unitflux covering by the approximate $7.0\times1.5\,\mathrm{arcmin}^2$ area by NGC 7090 is smaller than 0.04 \citep{moretti_etal03}, suggesting that X1 and X2 are unlikely to be background AGNs.

Most Galactic BHXRBs are transient X-ray sources with dramatic X-ray variability. If X1 and X2 are similar to BHXRBs, i.e. stellar massive BH with sub-Eddington accretion rate, then the mass of the BH should be around $30M_\odot$, assuming the observed peak luminosity are close to the Eddington luminosity (i.e. in the soft state, \citealt{remillard_mcclintock06}). However, the temperature ($2.07^{+0.30}_{-0.23}$ and $1.69^{+1.17}_{-0.48}$\,keV for X1 and X2, respectively) obtained from X-ray spectral analysis is inconsistent with the prediction for a disc around a $30M_\odot$ BH. The X-ray data of X1 have slightly better S/N, and can be fitted with a two component model. The \textsc{powerlaw+diskbb} model showed a hard ($1.44^{+0.26}_{-0.29}$) photon index with a weak and cool disc (the ratio of the disc flux to the total flux $f_\mathrm{disc}\sim0.19$, $T_\mathrm{in}=0.22^{+0.18}_{-0.08}$), which is consistent with the low/hard state of BHXRBs. If this is the case, the peak luminosity of X1 could be even higher, thus the BH mass should be much larger (e.g. an IMBH). But we note that the \textsc{powerlaw+diskbb} model does not improve the fit significantly comparing to the single component models. X1 also showed a transition in spectral shape with a much softer or cooler spectrum in 2006. This is reminiscent of the quiescent state ($\Gamma=1.5-2.1$, \citealt{remillard_mcclintock06}) in BHXRBs.

Alternatively, super-Eddington accretion onto an NS or a BH with mass less than $10M_\odot$ cannot be ruled out. High X-ray variability, though rare, has been found in some ULXs (e.g. \citealt{pintore_etal18} and references therein). All the four pulsating ULXs also showed high level flux variability. In a recent paper, \citet{walton_etal18b} showed that the broadband X-ray spectra of the bright ULXs can be fitted with a model consistent with super-Eddington accretion onto NSs, which may suggest that the compact object in many ULXs are neutron stars. The photon index and the disc temperature of X2, when fitted with the two simple models, are in agreement with the typical values found in ULXs with low S/N data (e.g. \citealt{makishima_etal00}) as well as the transient pulsar M82 X2 at a similar luminosity range \citep{brightman_etal16}. Similar to the other ULXs, the spectra of X1 in the high luminosity state can be described with a hot blackbody component plus a cool multicolour disc component. Though X1 did not show strong variability or pulsation during the 2007 \mission{XMM-Newton} observation, it is known that short-term variability and pulsation in some pulsar ULXs is transient. To further confirm the nature of the compact object of those two ULXs, future high S/N X-ray observations are necessary.

We did not find significant variability in $F814W$ flux for the X1 optical counterpart in the two \mission{HST} observations, which may suggest that the optical emission is from the companion star. Future simultaneous optical and X-ray observations are needed to confirm the nature of the companion star as well as the optical emission, however.

\section*{Acknowledgements}

ZL thanks the support from the China Scholarship Council. This work is supported by the Strategic Pioneer Program on Space Science, Chinese Academy of Sciences, Grant No. XDA15052100. PTOB acknowledges support from STFC. JPO, PAE and KLP acknowledge support from the UK Space Agency. This work made use of data supplied by the UK Swift Science Data Centre at the University of Leicester. This work is based on observation obtained with \mission{XMM-Newton}, an ESA science mission with instruments and contributions directly fund by ESA Member States and NASA, based on observations made with the NASA/ESA Hubble Space Telescope, and obtained from the Hubble Legacy Archive, which is a collaboration between the Space Telescope Science Institute (STScI/NASA), the Space Telescope European Coordinating Facility (ST-ECF/ESA) and the Canadian Astronomy Data Centre (CADC/NRC/CSA) The \mission{Chandra} data are obtained from the \mission{Chandra Data Archive}. This research has made use of software provided by the Chandra X-ray Center (CXC) in the application packages CIAO.

\bibliographystyle{mnras}
\bibliography{./references}

% Don't change these lines
\bsp	% typesetting comment
\label{lastpage}
\end{document}

%% file: obs_log.tex
\begin{table*}
\caption{\label{tab:obs_log}Observation logs}
\begin{tabular}{@{}llcccc@{}}\hline
Mission    & Observation/proposal ID & Observation date & Exposure time (s) &  Instrument/filters & Used in work \\\hline
\mission{XMM-Newton} & 0200230101 & 2004-04-08 &  27981  &  ---             &  N  \\
                     & 0200230201 & 2004-05-13 &  6024/10390/10290  & pn/M1/M2           &  Y  \\
                     & 0503460101 & 2007-10-05 &  4270/12790/12980  & pn/M1/M2           &  Y  \\[0.5mm]
\mission{Chandra}    & 7060       & 2005-12-18 &  26410  & ACIS-S           &  Y  \\
                     & 7252       & 2006-04-10 &  31020  & ACIS-S           &  Y  \\[0.5mm]
\mission{Swift}      & 35883001   & 2006-09-24 &  2349   & XRT              &  Y  \\
                     & 35883002   & 2006-10-31 &  2191   & XRT              &  Y  \\
                     & 32287001   & 2012-02-27 &  3856   & XRT              &  Y  \\
                     & 32287002   & 2012-03-12 &  3350   & XRT              &  Y  \\
                     & 32287003   & 2012-03-26 &  3161   & XRT              &  Y  \\
                     & 32287004   & 2012-04-09 &  3673   & XRT              &  Y  \\
                     & 32287005   & 2012-04-23 &  4091   & XRT              &  Y  \\
                     & 32287006   & 2012-05-07 &  4124   & XRT              &  Y  \\
                     & 32287007   & 2012-05-21 &  3952   & XRT              &  Y  \\
                     & 32287008   & 2012-06-04 &  2897   & XRT              &  Y  \\
                     & 32287009   & 2012-06-18 &  2681   & XRT              &  Y  \\
                     & 32287010   & 2012-06-21 &  1592   & XRT              &  Y  \\
                     & 32287011   & 2012-07-02 &  3852   & XRT              &  Y  \\
                     & 35883003   & 2012-07-24 &  2171   & XRT              &  Y  \\
                     & 32287012   & 2012-07-30 &  1972   & XRT              &  Y  \\
                     & 32287013   & 2012-08-02 &   602   & XRT              &  Y  \\
                     & 32287014   & 2012-08-02 &  2093   & XRT              &  Y  \\
                     & 32287015   & 2012-08-20 &  1953   & XRT              &  Y  \\
                     & 84548001   & 2014-06-15 &  1188   & XRT              &  Y  \\
                     & 84548002   & 2014-07-24 &   221   & XRT              &  Y  \\
                     & 84548003   & 2014-07-26 &   728   & XRT              &  Y  \\[0.5mm] 
\mission{HST}        & 5446       & 1994-06-20 &  160    & WFPC2/F606W      &  N  \\
                     & 9042       & 2001-09-24 &  460, 460     & WFPC2/F450W,F814W   &  Y  \\
                     & 10416      & 2005-06-23 &  2508, 7496   & ACS/F625W, F658N    &  N  \\
                     & 10889      & 2007-05-17 &  4000         & WFPC2/F814W         &  N  \\
                     & 12546      & 2012-04-09 &  900, 900     & ACS/F606W, F814W    &  Y  \\
                     & 14095      & 2016-03-08 &  298, 1802    & WFC3/F110W, F128N   &  N  \\\hline
\end{tabular}
\end{table*}

%% file: table_fitting.tex
\begin{table}
\setlength\tabcolsep{3.0pt}
\caption{\label{tab:fit_para}Best-fitting parameters}
\begin{tabular}{@{}llccccc@{}}\hline
    Model                  & $N_{\mathrm{H, host}}$ & $\Gamma$/$T_\mathrm{BB}$ & $T_{\mathrm{in}}$       & $\log f_{\mathrm{0.3-10 keV}}^b$    & $C_{\mathrm{stat}}/\mathrm{d.o.f}$ \\[1.0mm]
                       & $10^{22}\,\mathrm{cm}^{-2}$ &   $\mathrm{keV}$    &  $\mathrm{keV}$        & $\mathrm{erg\,s^{-1}\,cm^{-2}}$ &  \\\hline
\multicolumn{6}{c}{X1 \mission{XMM-Newton} observation (2007)}                              \\[1mm]
\textsc{po}            & $0.5^{+0.1}_{-0.1}$    & $1.55^{+0.15}_{-0.15}$   &     --                  & $-12.09^{+0.04}_{-0.04}$ & $708.67/815$  \\[1.2mm]
\textsc{diskbb}        & $0.3^{+0.1}_{-0.1}$    &     --                   & $2.07^{+0.30}_{-0.23}$  & $-12.19^{+0.05}_{-0.05}$ & $722.32/815$  \\[1.2mm]
\textsc{po+diskbb}     & $0.8^{+0.4}_{-0.3}$    & $1.44^{+0.26}_{-0.29}$   & $0.22^{+0.18}_{-0.08}$  & $-11.93^{+0.45}_{-0.19}$ & $703.23/813$  \\[1.2mm]
\textsc{bb+diskbb}     & $0.6^{+0.2}_{-0.2}$    & $1.41^{+0.24}_{-0.16}$   & $0.40^{+0.14}_{-0.09}$  & $-12.07^{+0.12}_{-0.09}$ & $697.34/813$  \\[2.0mm]
\multicolumn{6}{c}{X1 \mission{Chandra} observation (2006)}                              \\[1mm]
\textsc{po}            & $0.5^a$                & $2.67^{+0.69}_{-0.64}$   &     --                  & $-13.44^{+0.23}_{-0.17}$ & $36.47/28$    \\[1.2mm]
\textsc{diskbb}        & $0.3^a$                &     --                   & $0.64^{+0.28}_{-0.17}$  & $-13.78^{+0.12}_{-0.13}$ & $32.74/28$    \\[2.0mm]
\multicolumn{6}{c}{X2 \mission{Swift}1 observations (2012 June 04 to July 02)}                              \\[1mm]
\textsc{po}            & $0.3^{+0.4}_{-0.3}$    & $1.61^{+0.55}_{-0.50}$   &     --                  & $-12.12^{+0.10}_{-0.10}$ & $78.21/102$   \\[1.2mm]
\textsc{diskbb}        & $0.1^{+0.3}_{-0.1}$    &     --                   & $1.69^{+1.17}_{-0.48}$  & $-12.27^{+0.10}_{-0.10}$ & $77.43/102$   \\[2.0mm]
\multicolumn{6}{c}{X2 \mission{Swift}2 observations (2012 July 30 to August 20)}                              \\[1mm]
\textsc{po}            & $0.3^a$                & $1.59^{+1.04}_{-1.04}$   &     --                  & $-12.84^{+0.30}_{-0.30}$ & $12.86/16$    \\[1.2mm]
\textsc{diskbb}        & $0.1^a$                &     --                   & $>0.92$                 & $-12.87^{+0.38}_{-0.38}$ & $12.79/16$    \\\hline
\end{tabular}
\parbox[]{\columnwidth}{
    $a$: parameter is fixed.\\
    $b$: unabsorbed flux.
}
\end{table}